\documentclass[twocolumn,amssymb,amsmath,aps,showpacs,nofootinbib,nobibnotes,pra]{revtex4-1}

\usepackage{graphicx}
\usepackage{xcolor}
\usepackage[colorlinks=true,linkcolor=blue,urlcolor=blue,citecolor=blue,bookmarks=true,bookmarksopenlevel=2]{hyperref}
\usepackage{amsmath}

\usepackage[utf8]{inputenc}
\usepackage[T1]{fontenc}
\usepackage{microtype}
\usepackage{lmodern}

\usepackage{array,tabularx,booktabs}

\newcolumntype{L}{>{\raggedright\arraybackslash}X}
\newcolumntype{R}{>{\raggedleft\arraybackslash}X}
\newcolumntype{C}{>{\centering\arraybackslash}X}

\usepackage{rotating}

\newcommand{\ket}[1]{\left\vert#1\right\rangle}

\usepackage{xspace}
\newcommand{\WNi}{W$^{46+}$\xspace}
\newcommand{\WCu}{W$^{45+}$\xspace}

\usepackage{siunitx}

\usepackage[caption=false]{subfig}
\usepackage[normalem]{ulem}
\usepackage{soul}
\soulregister\cite7
\soulregister\ref7
\soulregister\pageref7

\begin{document}

\title{MCDHF-CI study of $4d-3p$ X-ray transitions in Cu- and Ni-like tungsten ions}
\author{Karol Kozio{\l}}
\email{Karol.Koziol@ncbj.gov.pl}
\author{Jacek Rzadkiewicz} 
\email{Jacek.Rzadkiewicz@ncbj.gov.pl}
\affiliation{Narodowe Centrum Bada\'{n} J\k{a}drowych (NCBJ), Andrzeja So{\l}tana 7, 05-400 Otwock-\'{S}wierk, Poland}

\begin{abstract}
The $4d \to 3p$ X-ray transitions in Cu- and Ni-like tungsten ions have been studied theoretically.
The Multiconfiguration Dirac--Hartree--Fock (MCDHF) method and the large-scale relativistic configuration interaction (CI) method have been employed in order to take into account electron correlation effects on the wavelengths and transition rates.
It was found that the wavelengths and transition rates obtained from the MCDHF-CI method depend strongly on the size and the type of the Active Space used in the calculations.
It has been found that extending the Active Space of orbitals without careful control of the Configuration State Function base does not always lead to good quality MCDHF-CI results for highly ionized tungsten ions.
\end{abstract}

\maketitle

\section{Introduction}

Tungsten has been chosen as a plasma facing material in modern large tokamaks, including JET and ITER \cite{Matthews2007,Bolt2002}.
Therefore, spectroscopic studies of tungsten ions can constitute a unique tool for diagnostics relevant for a wide range of electron temperatures, from 0.1~keV at the edge up to 25~keV in the core of the tokamak plasma \cite{Putterich2008}.
The tungsten spectra originating from different plasma regions with various electron temperatures consist of radiation emitted by many specific ion charges.
The spectra related to plasma temperatures above a few keV consist of many intense lines originating from Cu- (\WCu) and Ni-like (\WNi) tungsten ions.
These lines have been observed at the ASDEX Upgrade and JT60U tokamaks in various spectral ranges.
The $4d-3p$ x-ray lines of \WCu and \WNi ions were also observed at JET in the wavelength region of 5.19--5.24~\AA{} by means of an upgraded high-resolution x-ray diagnostic~\cite{Rzadkiewicz2013a,Chernyshova2014,Shumack2014,Nakano2015}.

In our previous paper \cite{Rzadkiewicz2018} the x-ray transitions in Ni- and Cu-like tungsten ions in the 5.19--5.26~\AA{} wavelength range that are relevant as a high-temperature tokamak diagnostic, in particular for JET in the  ITER-like wall configuration, have been studied.
The experimental wavelengths were measured at the upgraded Shanghai Electron Beam Ion Trap with an accuracy of 0.3--0.4~m\AA{}, and then compared with those determined from JET ITER-like wall plasmas.
It has been found that employing the Multiconfiguration Dirac--Hartree--Fock (MCDHF) calculations extended by taking into account correlation effects (Configuration Interaction, CI, approach) brings the calculations closer to the experimental values in comparison with previous calculations found in the literature.
In that paper, theoretical studies were presented only for the x-ray lines appearing in the measured spectra, i.e. for the so-called Ni1, Ni2, Cu1, Cu2, and Cu3 lines related to the transitions between excited $[Mg]3p^53d^{10}4d^1$/$[Mg]3p^53d^{10}4s^14d^1$ states and $[Mg]3p^63d^{10}$/$[Mg]3p^63d^{10}4s^1$ ground states, for \WNi and \WCu ions, respectively.

In the present paper we consider the wavelengths and intensities of all transitions between $\ket{[Mg]3p^53d^{10}4d^1}_{J=1}$ and $\ket{[Mg]3p^63d^{10}}_{J=0}$ states in Ni-like (\WNi) tungsten and between $\ket{[Mg]3p^53d^{10}4s^14d^1}_{J=1/2,3/2}$ and $\ket{[Mg]3p^63d^{10}4s^1}_{J=1/2}$ states in Cu-like (\WCu) tungsten.
In addition we present our methodology in more detail.
See Table~\ref{tab:state-label} for the terminology for the states.
Excited states may be divided into two groups: lower-lying states (states 1 and 2 for \WNi and states 1--7 for \WCu) and higher-lying states (state 3 for \WNi and states 8--11 for \WCu).
Transitions between lower-lying excited states and ground states for \WNi and \WCu ions appear in spectra in the range 5.19--5.30~\AA{}.
Transitions from higher-lying excited states appear in spectra in the range 4.6--4.7~\AA{}.

\begin{table}[!htb]
\caption{\label{tab:state-label}Description of states in Ni-like (\WNi) and Cu-like (\WCu) tungsten ions.}
\begin{ruledtabular}
\begin{tabular}{lll}
State & Term & $^{2S+1}L_J$\\
\midrule
\multicolumn{3}{@{}l}{\WNi}\\[1ex]
0 & $\ket{[Mg]3p^63d^{10}}_{J=0}$ & $^1S_{0}$\\[1ex]
1 & $\ket{[Mg]3p_{1/2}^2 3p_{3/2}^3 3d^{10} 4d_{3/2}^1}_{J=1}$ & $^3P_{1}$\\[1ex]
2 & $\ket{[Mg]3p_{1/2}^2 3p_{3/2}^3 3d^{10} 4d_{5/2}^1}_{J=1}$ & $^1P_{1}$\\[1ex]
3 & $\ket{[Mg]3p_{1/2}^1 3p_{3/2}^4 3d^{10} 4d_{3/2}^1}_{J=1}$ & $^3D_{1}$\\[1ex]
\multicolumn{3}{@{}l}{\WCu}\\[1ex]
0 & $\ket{[Mg]3p^63d^{10}4s^1}_{J=1/2}$ & $^2S_{1/2}$\\[1ex]
1 & $\ket{[Mg]\left[ 3p_{1/2}^2 3p_{3/2}^3 3d^{10} 4s^1 \right]_2 4d_{3/2}^1}_{J=1/2}$ & $^4P_{1/2}$\\[1ex]
2 & $\ket{[Mg]\left[ 3p_{1/2}^2 3p_{3/2}^3 3d^{10} 4s^1 \right]_2 4d_{3/2}^1}_{J=3/2}$ & $^4P_{3/2}$\\[1ex]
3 & $\ket{[Mg]\left[ 3p_{1/2}^2 3p_{3/2}^3 3d^{10} 4s^1 \right]_1 4d_{3/2}^1}_{J=1/2}$ & $^2P_{1/2}$\\[1ex]
4 & $\ket{[Mg]\left[ 3p_{1/2}^2 3p_{3/2}^3 3d^{10} 4s^1 \right]_1 4d_{3/2}^1}_{J=3/2}$ & $^2D_{3/2}$\\[1ex]
5 & $\ket{[Mg]\left[ 3p_{1/2}^2 3p_{3/2}^3 3d^{10} 4s^1 \right]_1 4d_{5/2}^1}_{J=3/2}$ & $^2P_{3/2}$\\[1ex]
6 & $\ket{[Mg]\left[ 3p_{1/2}^2 3p_{3/2}^3 3d^{10} 4s^1 \right]_2 4d_{5/2}^1}_{J=1/2}$ & $^2P_{1/2}$\\[1ex]
7 & $\ket{[Mg]\left[ 3p_{1/2}^2 3p_{3/2}^3 3d^{10} 4s^1 \right]_2 4d_{5/2}^1}_{J=3/2}$ & $^2D_{3/2}$\\[1ex]
8 & $\ket{[Mg]\left[ 3p_{1/2}^1 3p_{3/2}^4 3d^{10} 4s^1 \right]_0 4d_{3/2}^1}_{J=3/2}$ & $^4F_{3/2}$\\[1ex]
9 & $\ket{[Mg]\left[ 3p_{1/2}^1 3p_{3/2}^4 3d^{10} 4s^1 \right]_1 4d_{3/2}^1}_{J=1/2}$ & $^4D_{1/2}$\\[1ex]
10 & $\ket{[Mg]\left[ 3p_{1/2}^1 3p_{3/2}^4 3d^{10} 4s^1 \right]_1 4d_{3/2}^1}_{J=3/2}$ & $^4D_{3/2}$\\[1ex]
11 & $\ket{[Mg]\left[ 3p_{1/2}^1 3p_{3/2}^4 3d^{10} 4s^1 \right]_1 4d_{5/2}^1}_{J=3/2}$ & $^2D_{3/2}$\\[1ex]
\end{tabular}
\end{ruledtabular}
\end{table}

\section{Theoretical background}

The calculations of the radiative transition energies and rates have been carried out by means of the \textsc{Grasp2k} v1.1 \cite{Jonsson2007,Jonsson2013} code.
The \textsc{Grasp2k} code is based on the MCDHF method.
For comparison we have perform also calculations with \textsc{Fac} code based on the Dirac--Hartree--Fock--Slater method~\cite{Gu2008}.

The methodology of MCDHF calculations performed in the present study is similar to that published earlier in many papers (see, e.g., \cite{Dyall1989,Grant2007}).
The effective Hamiltonian for an $N$-electron system is expressed by
\begin{equation}
H = \sum_{i=1}^{N} h_{D}(i) + \sum_{j>i=1}^{N} C_{ij},
\end{equation}
where $h_D(i)$ is the Dirac operator for the $i$th electron and the terms $C_{ij}$ account for the electron--electron interactions.
In general, the latter is a sum of the Coulomb interaction operator and the transverse Breit operator.
An atomic state function (ASF) with total angular momentum $J$ and parity $p$ is assumed in the form
\begin{equation}
\Psi_{s} (J^{p} ) = \sum_{m} c_{m} (s) \Phi ( \gamma_{m} J^{p} ),
\end{equation}
where $\Phi ( \gamma_{m} J^{p} )$ are the configuration state functions (CSFs), $c_{m} (s)$ are the configuration mixing coefficients for state $s$, and $\gamma_{m}$ represents all information required to define a certain CSF uniquely.
The CSFs are linear combinations of $N$-electron Slater determinants which are antisymmetrized products of 4-component Dirac orbital spinors.
In present calculations, the initial and final states of the considered transitions have been optimized separately and a biorthonormal transformation has been used for performing the transition rate calculations \cite{Jonsson2007}.
Following this, the so-called relaxation effect is taken into account.

In the \textsc{Grasp2k} code, the Breit interaction contribution to the energy is added perturbatively, after the radial part of wavefunction has been optimized. 
We calculated the Breit term in low-frequency limit (see, e.g., \cite{Kozio2018} for details), because frequency-dependent term is not appropriate for virtual orbitals \cite{Si2018}. 
Using self-consistent variational approach to calculate Breit term instead of perturbational approach (see, e.g., \cite{Kozio2018} for details) cause small effect, estimated to be about $+0.3$ to $+0.4$~m\AA{} at multi-reference (MR) MCDHF level in studied cases. 
However, it has been found that this ''variational effect'' is significantly reduced when active space is expanding \cite{Si2018,Chantler2014}. 
Also two types of quantum electrodynamics (QED) corrections: the self-energy (as the screened hydrogenic approximation \cite{McKenzie1980} of the data of Mohr and co-workers \cite{Mohr1992a}) and the vacuum polarization (as the potential of Fullerton and Rinker \cite{Fullerton1976}) have been included. 
The differences in various models of estimating the self-energy in many-electron atoms (see, e.g., \cite{Kozio2018a} for details) cause effect below 0.1~m\AA{} in studied cases. 
The radiative transition rates were calculated in the Babushkin (length) \cite{Babushkin1964} gauge.

The accuracy of the wavefunction depends on the CSFs included in its expansion \cite{FroeseFischer2010,FroeseFischer2011}.
The accuracy can be improved by extending the CSF set by including the CSFs originating from excitations from orbitals occupied in the reference CSFs to unfilled orbitals of the active orbital set (i.e., CSFs for virtual excited states).
This approach is called Configuration Interaction. 
The CI method makes it possible to include the major part of the electron correlation contribution to the energy of the atomic levels.
In the CI approach, it is very important to choose an appropriate basis of CSFs for the virtual excited states.
It can be done by systematically building CSF sequences by extending the Active Space (AS) of orbitals and concurrently monitoring the convergence of the self-consistent calculations \cite{FroeseFischer2010,Lowe2010,Fei2012}.

\begin{table}[!htb]
\caption{\label{tab:cfs-no-ni}Numbers of CSFs for different active spaces used in calculations of considered transitions in Ni-like (\WNi) tungsten ions.}
\begin{ruledtabular}
\begin{tabular}{llrr}
& & \multicolumn{2}{c}{Number of CSFs}\\
\cmidrule{3-4}
Active & Virtual & upper & lower\\
space & orbitals & states & state\\\midrule	
AS0 & & 3 & 1 \\
AS1d & $n=4$, $l=0{-}2$ & 2741 & 159 \\
AS1f & $n=4$, $l=0{-}3$ & 8845 & 302 \\
AS2d & $n=4{-}5$, $l=0{-}2$ & 17102 & 593 \\
AS2f & $n=4{-}5$, $l=0{-}3$ & 44345 & 1147\\
AS2g & $n=4{-}5$, $l=0{-}4$ & 58344 & 1417 \\
AS3d & $n=4{-}6$, $l=0{-}2$ & 43513 & 1303 \\
AS3f & $n=4{-}6$, $l=0{-}3$ & 106930 & 2536\\
AS3g & $n=4{-}6$, $l=0{-}4$ & 152800 & 3370 \\
AS4d & $n=4{-}7$, $l=0{-}2$ & 81974 & 2289 \\
AS4f & $n=4{-}7$, $l=0{-}3$ & 196600 & 4469\\
AS4g & $n=4{-}7$, $l=0{-}4$ & 292213 & 6161 \\
\end{tabular}
\end{ruledtabular}
\end{table}

\begin{table}[!htb]
\caption{\label{tab:cfs-no-cu}Numbers of CSFs for different active spaces used in calculations of considered transitions in Cu-like (\WCu) tungsten ions -- CV approach.}
\begin{ruledtabular}
\begin{tabular}{llrr}
& & \multicolumn{2}{c}{Number of CSFs}\\
\cmidrule{3-4}
Active & Virtual & upper & lower\\
space & orbitals & states & state\\\midrule	
AS0 & & 11 & 1 \\
AS1 & $n=4$, $l=0{-}3$ & 4571 & 77 \\
AS2 & $n=4{-}5$, $l=0{-}4$ & 26594 & 362 \\
AS3 & $n=4{-}6$, $l=0{-}4$ & 68624 & 866 \\
AS4 & $n=4{-}7$, $l=0{-}4$ & 130661 & 1589\\
\end{tabular}
\end{ruledtabular}
\end{table}

\begin{table}[!htb]
\caption{\label{tab:cfs-no-ni2}Numbers of CSFs for different active spaces used in calculations of considered transitions in Ni-like (\WNi) tungsten ions -- CV approach.}
\begin{ruledtabular}
\begin{tabular}{llrr}
& & \multicolumn{2}{c}{Number of CSFs}\\
\cmidrule{3-4}
Active & Virtual & upper & lower\\
space & orbitals & states & state\\\midrule	
AS0 & & 3 & 1 \\
AS1 & $n=4$, $l=0{-}3$ & 5434 & 169 \\
AS2 & $n=4{-}5$, $l=0{-}4$ & 37036 & 832 \\
AS3 & $n=4{-}6$, $l=0{-}4$ & 96503 & 1998 \\
AS4 & $n=4{-}7$, $l=0{-}4$ & 183835 & 3667\\
\end{tabular}
\end{ruledtabular}
\end{table}

\begin{figure*}[!htb]
\subfloat[]{\includegraphics[width=0.49\columnwidth]{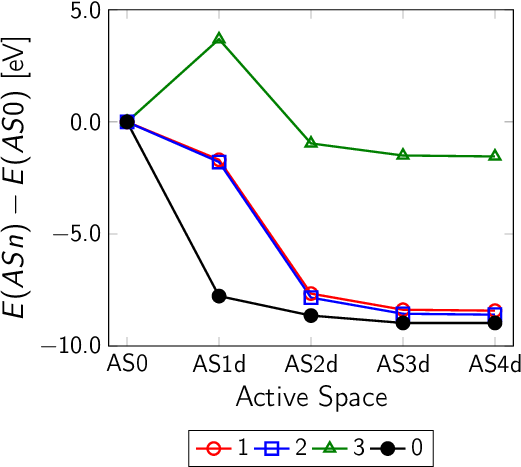}}\hfill
\subfloat[]{\includegraphics[width=0.49\columnwidth]{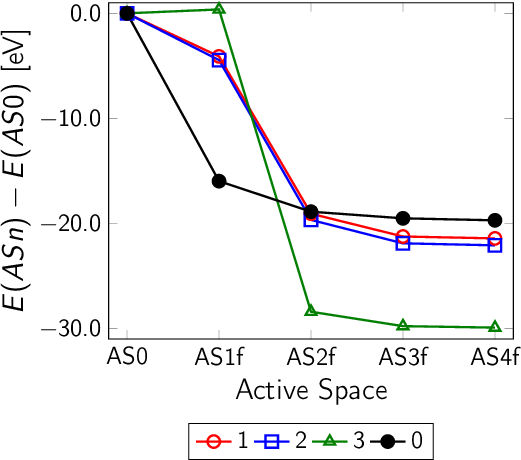}}\hfill
\subfloat[]{\includegraphics[width=0.49\columnwidth]{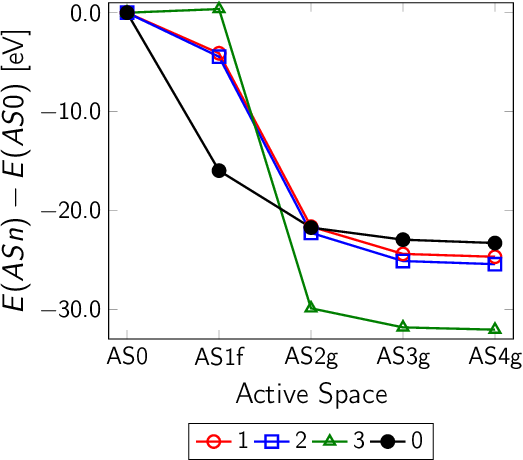}}\hfill
\subfloat[]{\includegraphics[width=0.49\columnwidth]{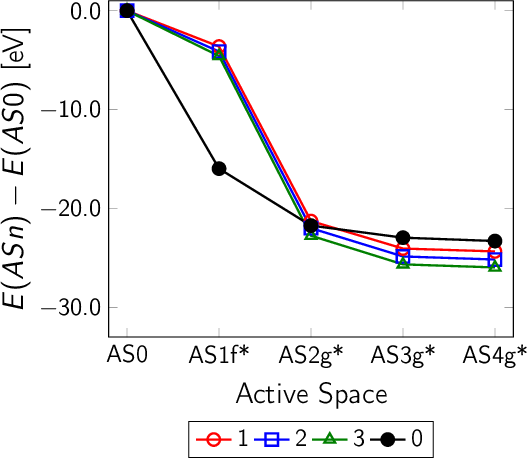}}
\caption{\label{fig:ni-lev}
Energies (in eV) of $\ket{[Mg]3p^53d^{10}4d^1}_{J=1}$ atomic levels (marked as 1, 2, 3 from lowest to highest) and $\ket{[Mg]3p^63d^{10}}_{J=0}$ level (marked as 0) for Ni-like tungsten ions calculated for various CI bases, relative to the ones calculated for the reference basis. 
(a) FCI-d, (b) FCI-f, (c) FCI-g, (d) FCI-g* model, respectively. 
}
\end{figure*}

\begin{figure}[!htb]
\subfloat[]{\includegraphics[width=0.49\columnwidth]{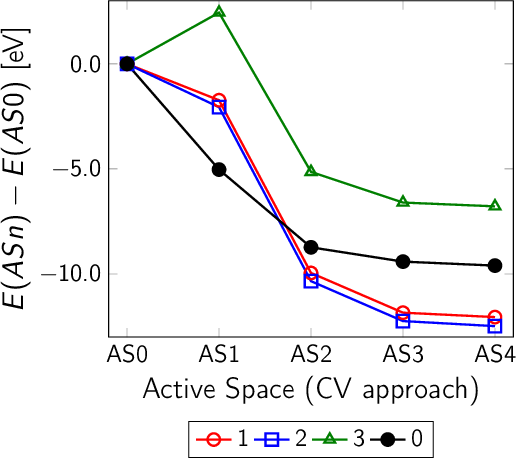}}\hfill
\subfloat[]{\includegraphics[width=0.49\columnwidth]{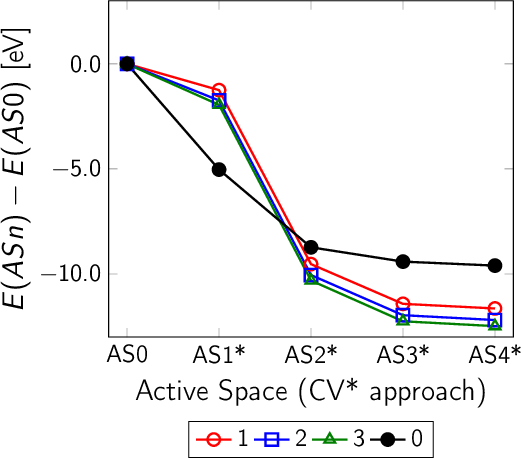}}
\caption{\label{fig:ni-lev2}
As in Figure~\ref{fig:ni-lev}, but for CV approaches: (a) CV and (b) CV* model, respectively.
}
\end{figure}

The $\ket{[Mg]3p^53d^{10}4s^{0,1}4d^1}\to\ket{[Mg]3p^63d^{10}4s^{0,1}}$ transitions  in Cu- and Ni-like W ions are interesting within the CI framework because: (i) the lower states of the transitions are the lowest states for the given number of electrons in the given symmetry, but the upper states of the transitions are not the lowest states in the given symmetry; (ii) for the lower states of the transitions, all the virtual orbitals are above the occupied ones, but for the upper states of the transitions, some of the virtual orbitals are above and others are below the occupied ones.
Hence, SCF variational procedure need to be performed carefully.

For \WNi we have tested the active spaces of virtual orbitals with $n$ up to $n=7$ and $l$ up to $l=5$.
Details are presented in Table~\ref{tab:cfs-no-ni}.
We have considered all possible single (S) and double (D) substitutions from the $3s$, $3p$, $3d$, $4d$ occupied subshells for the upper states and from the $3s$, $3p$, $3d$ occupied subshells for the lower state.
In this case, the inactive core contains $n = 1, 2$ subshells.
Some CSFs (the ones that have a zero matrix element with every CSF in the MR set) are excluded by using the \textit{jjreduce3} program, a part of the \textsc{Grasp2k} program set.
In this way the number of CSFs has been reduced by up to 35\%.
The MR configurations are $[Mg]3p^53d^{10}4d^1$ and $[Mg]3p^63d^{10}$ for the upper and lower states of the mentioned transitions in \WNi ions (giving AS0 active space).
This approach will be referred to as the FCI (``full CI'').

Because the size of the expansions increases rapidly with the size of the reference set, for \WCu we used a slightly simplified model, which is a common approach (see e.g., \cite{Fei2012,FroeseFischer2009a}).
The occupied subshells have been divided into three kinds: inactive core, active core (C), and valence (V) subshells.
The $n = 4$ subshells (i.e., $4s$ and $4d$ for the upper states of transitions and $4s$ for the lower state) are considered as valence subshells.
The $n = 1, 2$ subshells are an inactive core and the $n = 3$ subshells are an active core.
Then, for \WCu we considered SD substitutions divided into two groups: VV (both substituted electrons are from the valence subshells) and CV (the first substituted electron is from the valence subshell and the other is from the active core subshell) substitutions.
The details are presented in Table~\ref{tab:cfs-no-cu}.
The reference (AS0) configurations are $[Mg]3p^53d^{10}4s^14d^1$ and $[Mg]3p^63d^{10}4s^1$ for the upper and lower states of mentioned transitions in \WCu ion.
This approach will be referred to as CV (``core--valence'').

In order to prove that the CV approach provides results as good as the more sophisticated FCI approach, we used the CV approach also in the case of Ni-like tungsten ions.
In this case the open subshells ($3p$ and $4d$ for the upper states but nothing for the lower state) are treated as valence subshells.
The $3s$ and $3d$ subshells for the upper states and the $3s$, $3p$, and $3d$ ones for the lower state are an active core.
The $n = 1, 2$ subshells are an inactive core. 
VV and CV SD substitutions are allowed.
The details are presented in Table~\ref{tab:cfs-no-ni2}.

\section{Results and discussion}

Firstly, we studied the convergence of the energies of the initial and final states in the CI procedure.
We considered three FCI models: FCI-d (Fig.~\ref{fig:ni-lev}a), when virtual orbitals with maximal $l=2$ are used, FCI-f (Fig.~\ref{fig:ni-lev}b), when virtual orbitals with maximal $l=3$ are used, and FCI-g (Fig.~\ref{fig:ni-lev}c), when virtual orbitals with maximal $l=4$ are used.
As one can see, for all FCI models, the energy saturation for the ground state of Ni-like tungsten (0 level) precedes the saturation for the excited states (1, 2, 3 levels).
One can conclude that using the AS4g active space is high enough to reach the saturation at a similar level of relative energy for each initial and final state on the $n$ and $l$ levels.
The initial and final states of transitions in \WCu and \WNi have fairly different occupation numbers in the outer and inner shells.
Therefore, for the AS with low $n$ and $l$ the correlation corrections to the energy levels are very unequal for the initial and final states of the transitions.
This is due to the fact that some important substitutions from the $4d$ subshell are not allowed when the AS contains only subshells with $l \le 2$.
In other words, the energy of the final (ground) state converges for lower $n$ in AS in comparison with the energy for the initial states.
This inequality decreases for ASs with higher $n$ and, as a result, the values of the wavelength saturate at AS3--AS4.

In the case of the FCI-d model (see Fig.~\ref{fig:ni-lev}) a reduction of the energy at AS4d is larger for the ground state (level 0) than for the excited states 1 and 2, while state 3 nearly does not change its energy.
The situation is fundamentally different in the FCI-f and FCI-g models.
The reduction of energy for state 0 is slightly smaller than for states 1 and 2, and the reduction of energy for state 3 is distinctly larger.
The differences of energy of states 0, 1, 2, and 3 translates directly into the energies of the $1\to0$, $2\to0$, and $3\to0$ transitions (Ni2, Ni1, and Ni3 lines, respectively), calculated in various FCI models.
A similar analysis of the convergence for the CV approach is presented in Fig.~{\ref{fig:ni-lev2}a}.
In this case the reduction of energy for state~0 is smaller than for states~1 and~2 by a few eV and larger than for state 3, also by a few eV.

\begin{table}[!htb]
\caption{\label{tab:ni123en}Wavelengths of Ni1, Ni2, and Ni3 transitions (\AA) for various theoretical approaches.}
\begin{ruledtabular}
\begin{tabular*}{\linewidth}{@{\extracolsep{\fill}}llll}
& {Ni1} & {Ni2} & {Ni3}\\
& 0$\to$2 & 0$\to$1 & 0$\to$3\\
\midrule
\multicolumn{4}{l}{theory:}\\
MCDHF (MR) & 5.1947\footnotemark[1] & 5.2486\footnotemark[1] & 4.6367\\
MCDHF-CI (FCI-d) & 5.1939 & 5.2474 & 4.6238\\
MCDHF-CI (FCI-f) & 5.1999 & 5.2525 & 4.6544\\
MCDHF-CI (FCI-g) & 5.1994\footnotemark[1] & 5.2517\footnotemark[1] & 4.6519\\
MCDHF-CI (FCI-g*) & 5.1988 & 5.2510 & 4.6413\\
MCDHF-CI (CV) & 5.2010\footnotemark[2] & 5.2541\footnotemark[2] & 4.6318\\
MCDHF-CI (CV*) & 5.2004 & 5.2532 & 4.6417\footnotemark[2]\\
\midrule
\multicolumn{4}{l}{experiments:}\\
EBIT \cite{Rzadkiewicz2018} & 5.2008(3) & 5.2540(3) & \\
JET \cite{Rzadkiewicz2018} & 5.2005(9) &  & \\
\midrule
\multicolumn{4}{l}{other theory:}\\
\textsc{Fac} (this paper) & 5.1959\footnotemark[1] & 5.2500\footnotemark[1] & 4.6370\\
MCDF \cite{Dong2003} & 5.1942 & 5.2472 & 4.6265\\
MBPT \cite{Safronova2006} & 5.201 & 5.255 & 4.679\\
\textsc{Fac} \cite{Clementson2014} & 5.1963 & 5.2496 & 4.6287\\
\textsc{Relac}  \cite{Fournier1998} & 5.1994 &  & \\
\textsc{Cowan} \cite{Neill2004} & 5.218 & 5.272 & \\
\midrule
\multicolumn{4}{l}{other experiments:}\\
Ref. \cite{Clementson2010a} & 5.2002(9) & 5.2520(16) & 4.6372(10)\\
Ref. \cite{Tragin1988} & 5.203(3) & 5.255(3) & 4.638(3)\\
NIST \cite{Neill2004,Kramida2011} & 5.2004(9) & 5.2533(9) & \\
\end{tabular*}
\end{ruledtabular}
\footnotetext[1]{Numbers published in Rzadkiewicz \textit{et al.} \cite{Rzadkiewicz2018}}
\footnotetext[2]{The recommended values}
\end{table}

\begin{table}[!htb]
\caption{\label{tab:ni123tr}Transition rates of Ni1, Ni2, and Ni3 transitions  (s$^{-1}$, length gauge) for various theoretical approaches. Notation $A[B]$ means $A\times10^{B}$. The theoretical uncertainties are in parentheses (see text for details).}
\begin{ruledtabular}
\begin{tabular*}{\linewidth}{@{\extracolsep{\fill}}llll}
& {Ni1} & {Ni2} & {Ni3}\\
& 0$\to$2 & 0$\to$1 & 0$\to$3\\
\midrule
\multicolumn{4}{l}{theory:}\\
MCDHF (MR) & 1.05(2)[14] & 4.15(10)[12] & 4.84(10)[13]\\
MCDHF-CI (FCI-d) & 1.04(1)[14] & 6.24(4)[12] & 1.05(1)[13]\\
MCDHF-CI (FCI-f) & 9.22(8)[13] & 6.66(5)[12] & 6.02(3)[13]\\
MCDHF-CI (FCI-g) & 9.19(7)[13] & 7.14(7)[12] & 6.13(2)[13]\\
MCDHF-CI (FCI-g*) & 9.05(5)[13] & 6.63(4)[12] & 4.33(2)[13]\\
MCDHF-CI (CV) & 9.28(4)[13]\footnotemark[1] & 5.76(2)[12]\footnotemark[1] & 1.02(4)[13]\\
MCDHF-CI (CV*) & 9.13(6)[13] & 5.42(3)[12] & 4.31(4)[13]\footnotemark[1]\\
\midrule
\multicolumn{4}{l}{other theory:}\\
\textsc{Fac} (this paper) & 1.07[14] & 4.21[12] & 4.82[13]\\
MBPT \cite{Safronova2006} & 9.35[13] & 5.22[12] & 4.90[13]\\
\textsc{Fac} \cite{Clementson2014} & 9.52[13] &  & 1.68[13]\\
\textsc{Relac}  \cite{Fournier1998} & 9.56[13] &  & \\
\textsc{Cowan} \cite{Neill2004} & 1.25[14] & 8.85[12] & \\
\end{tabular*}
\end{ruledtabular}
\footnotetext[1]{The recommended values}
\end{table}

\begin{figure}[!htb]
\subfloat[]{\includegraphics[width=0.49\columnwidth]{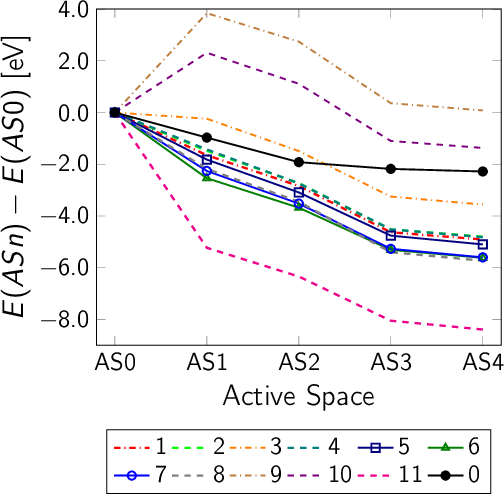}}\hfill
\subfloat[]{\includegraphics[width=0.49\columnwidth]{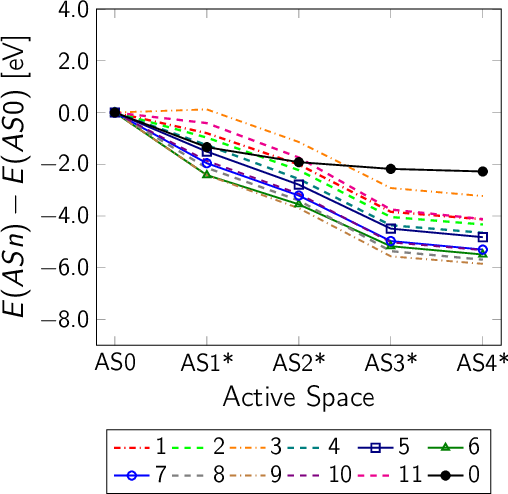}}
\caption{\label{fig:cu-lev}
Energies (in eV) of $\ket{[Mg]3p^53d^{10}4s^14d^1}_{J=1/2,3/2}$ atomic levels (marked as 1 to 11 from lowest to highest) and $\ket{[Mg]3p^63d^{10}4s^1}_{J=1/2}$ level (marked as 0) for Cu-like tungsten ions calculated for various CI bases, relative to the ones calculated for the reference basis.
Solid lines with markers are related to the states involved in the Cu1, Cu2, and Cu3 transitions. 
(a) CV and (b) CV* model, respectively.
}
\end{figure}

\begin{sidewaystable}
\caption{\label{tab:cuxen}Wavelengths of $\ket{[Mg]3p^53d^{10}4s^14d^1}_{J=1/2,3/2}\to\ket{[Mg]3p^63d^{10}4s^1}_{J=1/2}$ transitions (\AA) in Cu-like tungsten ions for various theoretical approaches.}
\begin{ruledtabular}
\begin{tabular*}{\linewidth}{@{\extracolsep{\fill}}l lllllllllll}
& 0$\to$1 & 0$\to$2 & 0$\to$3 & 0$\to$4 & 0$\to$5 & 0$\to$6 & 0$\to$7 & 0$\to$8 & 0$\to$9 & 0$\to$10 & 0$\to$11 \\
&  &  &  &  & {Cu3} & {Cu2} & {Cu1} &  &  &  &  \\
\midrule
\multicolumn{12}{l}{theory:}\\
MCDHF (MR) & 5.2968 & 5.2878 & 5.2760 & 5.2712 & 5.2308\footnotemark[1] & 5.2207\footnotemark[1] & 5.2178\footnotemark[1] & 4.6743 & 4.6685 & 4.6640 & 4.6192 \\
MCDHF-CI (CV) & 5.3028\footnotemark[2] & 5.2935\footnotemark[2] & 5.2789\footnotemark[2] & 5.2769\footnotemark[2] & 5.2370\footnotemark[1]\footnotemark[2] & 5.2281\footnotemark[1]\footnotemark[2] & 5.2251\footnotemark[1]\footnotemark[2] & 4.6804 & 4.6644 & 4.6624 & 4.6297 \\
MCDHF-CI (CV*) & 5.3010 & 5.2925 & 5.2781 & 5.2765 & 5.2364 & 5.2278 & 5.2245 & 4.6803\footnotemark[2] & 4.6748\footnotemark[2] & 4.6693\footnotemark[2] & 4.6223\footnotemark[2] \\
\multicolumn{12}{l}{experiments:}\\
EBIT \cite{Rzadkiewicz2018} &  &  &  &  & 5.2369(3) & 5.2292(3) & 5.2259(4) &  &  &  &  \\
JET \cite{Rzadkiewicz2018} &  &  &  &  &  & 5.2295(9) & 5.2263(9) &  &  &  &  \\
\midrule
\multicolumn{12}{l}{other theory:}\\
\textsc{Fac} (this paper) & 5.2979 & 5.2889 & 5.2770 & 5.2722 & 5.2316\footnotemark[1] & 5.2218\footnotemark[1] & 5.2191\footnotemark[1] & 4.6745 & 4.6686 & 4.6641 & 4.6195 \\
\textsc{Fac} \cite{Clementson2014} &  &  & 5.2728 &  & 5.2313 & 5.2230 & 5.2197 &  &  & 4.6559 &  \\
\textsc{Relac}  \cite{Fournier1998} &  &  &  &  & 5.2298 & 5.2192 &  &  &  & 4.6633 &  \\
\textsc{Cowan} \cite{Neill2004} &  &  &  &  & 5.241 & 5.230 &  &  &  &  &  \\
MBPT \cite{Safronova2003,Safronova2012} &  &  &  &  & 5.2409 & 5.2328 &  &  &  & 4.6601 & 4.6181 \\
\midrule
\multicolumn{12}{l}{other experiments:}\\
Ref. \cite{Osborne2011} &  &  &  &  & 5.238(9) &  &  &  &  &  &  \\
NIST \cite{Neill2004,Kramida2011} &  &  &  &  & 5.2379(17) & 5.2289(11) &  &  &  &  &  \\
\end{tabular*}
\end{ruledtabular}
\footnotetext[1]{Numbers published in Rzadkiewicz \textit{et al.} \cite{Rzadkiewicz2018}}
\footnotetext[2]{The recommended values}
\end{sidewaystable}

\begin{sidewaystable}
\caption{\label{tab:cuxrat}Transition rates of $\ket{[Mg]3p^53d^{10}4s^14d^1}_{J=1/2,3/2}\to\ket{[Mg]3p^63d^{10}4s^1}_{J=1/2}$ transitions (s$^{-1}$, length gauge) in Cu-like tungsten ions for various theoretical approaches. Notation $A[B]$ means $A\times10^{B}$. The theoretical uncertainties are in parentheses (see text for details).}
\begin{ruledtabular}
\begin{tabular*}{\linewidth}{@{\extracolsep{\fill}}l lllllllllll}
& 0$\to$1 & 0$\to$2 & 0$\to$3 & 0$\to$4 & 0$\to$5 & 0$\to$6 & 0$\to$7 & 0$\to$8 & 0$\to$9 & 0$\to$10 & 0$\to$11 \\
&  &  &  &  & {Cu3} & {Cu2} & {Cu1} &  &  &  &  \\
\midrule
\multicolumn{12}{l}{theory:}\\
MCDHF (MR)& 3.13(13)[11] & 2.03(6)[12] & 1.55(2)[13] & 4.78(33)[10] & 9.08(16)[13] & 9.11(18)[13] & 1.44(3)[13] & 3.12(13)[09] & 4.73(9)[13] & 4.46(9)[13] & 2.40(7)[12] \\
MCDHF-CI (CV) & 2.47(16)[11]\footnotemark[1] & 1.75(6)[12]\footnotemark[1] & 2.22(3)[13]\footnotemark[1] & 9.95(35)[10]\footnotemark[1] & 8.12(14)[13]\footnotemark[1] & 6.67(13)[13]\footnotemark[1] & 1.30(3)[13]\footnotemark[1] & 3.81(19)[10] & 6.45(36)[12] & 1.39(5)[13] & 6.04(41)[11] \\
MCDHF-CI (CV*) & 2.74(20)[11] & 1.64(7)[12] & 2.14(4)[13] & 8.07(53)[10] & 7.96(17)[13] & 6.55(15)[13] & 1.27(4)[13] & 3.00(41)[09]\footnotemark[1] & 4.19(8)[13]\footnotemark[1] & 3.71(8)[13]\footnotemark[1] & 2.22(10)[12]\footnotemark[1] \\
\midrule
\multicolumn{12}{l}{other theory:}\\
\textsc{Fac} (this paper) & 3.25[11] & 2.07[12] & 1.56[13] & 4.41[10] & 9.21[13] & 9.24[13] & 1.44[13] & 1.07[09] & 4.71[13] & 4.44[13] & 2.35[12] \\
\textsc{Fac} \cite{Clementson2014} &  &  & 2.21[13] &  & 8.14[13] & 6.58[13] & 1.31[13] &  &  & 1.32[13] &  \\
\textsc{Relac}  \cite{Fournier1998} &  &  &  &  & 8.32[13] & 7.80[13] &  &  &  & 3.99[13] &  \\
MBPT \cite{Safronova2003,Safronova2012} &  &  &  &  & 8.03[13] & 6.44[13] &  &  &  & 6.35[13] & 2.43[12] \\
\end{tabular*}
\end{ruledtabular}
\footnotetext[1]{The recommended values}
\end{sidewaystable}

The wavelengths and transition rates of the Ni1, Ni2, and Ni3 transitions, calculated by using various theoretical approaches, are presented in Tables \ref{tab:ni123en} and \ref{tab:ni123tr}.
They are compared with the experimental results presented in our previous paper \cite{Rzadkiewicz2018}, and also compared with experimental and other theoretical data available in the literature. 
On can see from Table \ref{tab:ni123en} that both the FCI-g and CV approaches are close to the experimental data for the N1 and Ni2 lines cited in \cite{Rzadkiewicz2018}, surrounding the experimental values from above and below.
The MCDHF-CI values are better than those of MR MCDHF or \textsc{Fac}.
However, in the case of the Ni3 line, the results of FCI-g and CV differ from each other and from the experimental values.
A disagreement between other theoretical attempts to predict the Ni3 wavelength can also be observed.
A similar disagreement is observed for the theoretical predictions for transition rate of Ni3 transition, see Table \ref{tab:ni123tr}.
There is a relatively large difference between the results from MCDHF-CI (FCI-g) and MCDHF-CI (CV), and a noticable difference between the results of MCDHF (MR) and MCDHF-CI (CV).
The latter is important, because the MCDHF-CI approach should be treated only as a correction to the MCDHF (MR) approach, so for strong transitions it is expected that the MCDHF-CI value is not substantially different from the MCDHF value.

Studying the CSF expansion of the ASF for level 3, one can see that this ASF is strongly influenced by the CSFs related to the $\ket{[Ne]3s^13p^63d^{10}4p^1}_{J=1}$ electronic configuration.
This is the result of an accidentally close energetic neighborhood of the virtual $\ket{[Ne]3s^13p^63d^{10}4p^1}_{J=1}$ level, originating from excitations to a virtual orbital, and level 3 of \WNi (the so called near-degeneracy effect, because $\ket{[Ne]3s^13p^63d^{10}4p^1}_{J=1}$ level formally is an excitation but actually has a lower energy than level 3).
Then we modified the FCI-g and CV approaches, removing two CSFs due to the $\ket{[Ne]3s^13p^63d^{10}4p^1}_{J=1}$ configuration from ASs. 
The new ones are referred to as the FCI-g* and CV* approaches.
One can see from Table \ref{tab:ni123en} that this modification changes the wavelengths of the Ni1 and Ni2 lines by less than 1~m\AA{}.
However, the change of wavelength for the Ni3 line is substantial, and the predictions calculated by the FCI-g* and CV* models met each other.
The similar consistency of theoretical predictions can be obtained in the case of the transition rates for the Ni3 transition -- see Table \ref{tab:ni123tr}.
It is worth noticing that the convergence of the atomic level energies from extending the AS is very similar for the FCI-g* and CV* approaches -- see Figs.~\ref{fig:ni-lev}d and~\ref{fig:ni-lev2}b.

The wavelengths and transition rates calculated by means of the various theoretical approaches for \WCu ions are presented in Tables~\ref{tab:cuxen} and~\ref{tab:cuxrat}, respectively.
The CV* represents the model when six CSFs due to the $[Ne]3s^13p^63d^{10}4s^14p^1$ configuration and one CSF due to the $[Ne]3s^2 3p_{1/2}^1 3p_{3/2}^4 3d^{10} 4p_{3/2}^2$ configuration are removed from the ASs.
The reason for applying this approach is similar to that for the CV* model in the case of the \WNi transitions: the $\ket{[Ne]3s^13p^63d^{10}4s^14p^1}_{J=1/2,3/2}$ and $\ket{[Ne]3s^2 3p_{1/2}^1 3p_{3/2}^4 3d^{10} 4p_{3/2}^2}$ levels are very close energetically to higher $\ket{[Ne]3s^23p^53d^{10}4s^14d^1}_{J=1/2,3/2}$ levels.
Removing these CSFs from the ASs improves the convergence of the atomic level energies -- compare Figs.~\ref{fig:cu-lev}a and~\ref{fig:cu-lev}b.
The difference between the results from the CV and the CV* is small for the transitions from lower-lying excited states, but significantly higher for the transitions from higher-lying excited states.

It is interesting to see that the transition rate for the strong Cu2 transition decreases markedly between the pure MCDHF (AS0) and the MCDHF-CI (AS4) frameworks.
This behavior may be explained by expanding the ASF for the initial state of the Cu2 transition in the LS-coupling CSF basis, and applying the rule for the selection of LS-coupling ($\Delta S = 0$; $\Delta L = 0, \pm 1$, but not $0\to0$), keeping in mind that the final state is $^2S$.
In the case of AS0, the initial state of the Cu2 transition is composed of 86.3\% $^2P$, 8.8\% $^4D$, and other CSFs.
In the case of AS4 it is composed of 72.7\% $^2P$, 7.5\% $^4D$, 7.3\% $^2S$ (related to excitation to virtual orbitals), and other CSFs.
In this case the $^2S$ term causes a reduction of the $^2P$ contribution, and at the same time the ${}^2S \to {}^2S$ transition is not allowed.
As a result, in the AS4 case the Cu2 transition rate is smaller than in the AS0 case.

Present work presents the correlation corrections on wavelengths and transition rates in various MCDHF-CI models. 
The ''correlation effect'' on transition wavelengths is estimated to be about 4--7~m\AA{} in studied cases and this effect is more significant for transition rates, which values may differ by a factor of a few depending on the model used in calculations. 
Basing on our best assessment, for analyzing future experiments we recommend MCDHF-CI (CV) values for Ni1, Ni2 lines in \WNi and $0 \to i$ ($i$~=~1--7) lines in \WCu, and MCDHF-CI (CV*) values for Ni3 line in \WNi and $0 \to i$ ($i$~=~8--11) lines in \WCu. 
The ''best'' numbers are chosen by comparing with experiments (when possible) and by carefully monitoring the convergence within MCDHF-CI process. 
The recommended values are marked in Tables \ref{tab:ni123en}, \ref{tab:ni123tr}, \ref{tab:cuxen}, and \ref{tab:cuxrat}. 
The theoretical uncertainties of wavelength numbers are related to convergence with the size of a basis set and estimated as absolute value of difference between wavelengths calculated within converged values ($AS\infty$; asymptote value assuming that correlation effects on energy levels are saturated, i.e. $|E(ASn{+}1)-E(ASn)|\to0$ when $n\to\infty$) and AS4 for given model, i.e. $\delta\lambda = |\lambda^{AS\infty}-\lambda^{AS4}|$. 
Such an estimation gives 0.1~m\AA{} uncertainty limit for all \WCu and \WNi theoretical wavelengths. 
More conservative estimation $\delta\lambda = |\lambda^{AS4}-\lambda^{AS3}|$ gives 0.1~m\AA{} uncertainty for \WNi lines and 0.3--0.5~m\AA{} for \WCu lines. 
In the case of transition rates, the total uncertainty contains also uncertainty related to difference between rates calculated by Babushkin (length) and Coulomb (velocity) gauges, i.e. $\delta A = \left[(A_{len}^{AS4}-A_{len}^{AS3})^2+(A_{len}^{AS4}-A_{vel}^{AS4})^2\right]^{1/2}$. 
For MR values, the uncertainty is related only to gauge difference, i.e. $\delta A = |A_{len}^{AS0}-A_{vel}^{AS0}|$, so comparing these uncertainties to other uncertainty values is not fully justified. 
The total uncertainties in transition rates are presented explicit in Tables \ref{tab:ni123tr} and \ref{tab:cuxrat}, being a test on the expansion model used. For recommended values they are below 1\% for \WNi lines and for \WCu lines they are below 3\% for strong lines ($A>10^{13}$~s$^{-1}$) and below 14\% for the week lines. 
It is worth to notice that in general low uncertainties indicate a higher quality of the theoretical model. 
For Ni1 and Ni2 lines the uncertainty for the recommended values are the smallest among considered models. 
For \WCu for lines $0 \to i$ ($i$~=~1--7) the CV model produces smaller uncertainties than CV* model, that proves our choice of recommended values. 
For Ni3 line and for $0 \to i$ ($i$~=~8--11) lines in \WCu transition rates vary a lot from one model to another. However, it is worth to notice that for these lines (except $0 \to 8$ line in \WCu which need more theoretical effort) the relative (percentage) uncertainty $\delta A/A$ is smaller for CV* model (recommended) than for CV model.

\section{Conclusions}

The energy levels of the ground excited states of \WCu and \WNi ions and the wavelengths and transition probabilities of the $4d \to 3p$ transitions were calculated by using the MCDHF-CI method.
It has been found that both MCDHF-CI approaches, FCI and CV, can provide reliable results if the active space is chosen properly.
It was found that if the highest occupied orbital in the initial or the final state of transition is the $nl$ orbital, then the active space should be extended to at least $(n+1)(l+1)$ virtual orbitals. 
Our analysis indicates that the best theoretical predictions of wavelengths and transition rates in the considered spectral range can be obtained by the MCDHF-CI (CV) model with the active spaces of virtual orbitals with $n$ up to $n$~=~7 and $l$ up to $l$~=~$5$. 
For higher-lying states (state 3 for \WNi and states 8--11 for \WCu) our recommended values of wavelengths and transition rates have been obtained the by CV approach with the removed specific CSFs in order to reduce the negative impact of the near-degeneration effect on the convergence in active spaces. 
This clearly shows that extending the active space of orbitals without careful control of the CSF basis does not always lead to good quality MCDHF-CI results for highly ionized tungsten ions. 
Results of present study provide an important benchmark for x-ray measurements in tokamaks, in particular for JET and ITER.


%

\end{document}